# Efficient Formulation of Polarizable Gaussian Multipole Electrostatics for Biomolecular Simulations

Haixin Wei[†], Ruxi Qi[†], Junmei Wang[‡], Piotr Cieplak[⊥], Yong Duan[§], and Ray Luo[†,*]

[†]Departments of Molecular Biology and Biochemistry, Chemical and Biomolecular Engineering, Materials Science and Engineering, and Biomedical Engineering, Graduate Program in Chemical and Materials Physics, University of California, Irvine, Irvine, California 92697, United States

[‡] Department of Pharmaceutical Sciences and Computational Chemical Genomics Screening Center, School of Pharmacy, University of Pittsburgh, Pittsburgh, Pennsylvania 15261, United States

[⊥]SBP Medical Discovery Institute, 10901 North Torrey Pines Road, La Jolla, California 92037, United States

[§]UC Davis Genome Center and Department of Biomedical Engineering, University of California, Davis, One Shields Avenue, Davis, California 95616, United States

Corresponding authors: ray.luo@uci.edu

## Abstract

Molecular dynamics simulations of biomolecules have been widely adopted in biomedical studies. As classical point-charge models continue to be used in routine biomolecular applications, there have been growing demands on developing polarizable force fields for handling more complicated biomolecular processes. Here we focus on a recently proposed polarizable Gaussian Multipole (pGM) model for biomolecular simulations. A key benefit of pGM is its screening of all short-range electrostatic interactions in a physically consistent manner, which is critical for stable charge-fitting and is needed to reproduce molecular anisotropy. Another advantage of pGM is that each atom's multipoles are represented by a single Gaussian function or its derivatives, allowing for more efficient electrostatics than other Gaussian-based models. In this study we present an efficient formulation for the pGM model defined with respect to a local frame formed with a set of covalent basis vectors. The covalent basis vectors are chosen to be along each atom's covalent bonding directions. The new local frame allows molecular flexibility during molecular simulations and facilitates an efficient formulation of analytical electrostatic forces without explicit torque computation. Subsequent numerical tests show that analytical atomic forces agree excellently with numerical finite-difference forces for the tested system. Finally, the new pGM electrostatics algorithm is interfaced with the PME implementation in Amber for molecular simulations under the periodic boundary conditions. To validate the overall pGM/PME electrostatics, we conducted an NVE simulation for a small water box of 512 water molecules. Our results show that, to achieve energy conservation in the polarizable model, it is important to ensure enough accuracy on both PME and induction iteration. It is hoped that the reformulated pGM model will facilitate the development of future force fields based on the pGM electrostatics for applications in biomolecular systems and processes where polarization plays crucial roles.

# 1. Introduction

Atomistic simulations of biomolecules have been applied in a wide range of biological systems.[1] While additive nonpolarizable models will continue to play important roles,[2-4] nonadditive polarizable models are expected to extend our ability to study more complex biomolecular systems and processes. Nonpolarizable models typically use fixed atom-centered partial charges to model electrostatics and include polarization response to the environment (mostly in water) only in an averaged, mean-field manner. Subsequently, nonpolarizable models that provide excellent descriptions of the homogeneous bulk phase are poor models for gas-phase clusters or in nonpolar solvents. The importance of modeling nonadditive effects is well known.[5] For example, the gas-phase water dimer interaction energy is overestimated by more than 30% in the TIP5P model.[6] Similarly for large biomolecular systems, there are concerns that such models cannot correctly account for situations where the same nonpolarizable moiety is exposed to different electrostatic environments/solvents, either within a single large structure or during a simulation process. In addition, there is an inherent inconsistency in most nonpolarizable models related to their static inclusion of average bulk polarization within the potential. This results in internal energies and other properties that are derived against a gas-phase reference state, which is already "pre-polarized" for the liquid phase. These limitations lead to issues in modeling multiple important problems such as pH-dependent processes, ion-dependent interactions, order-disorder transition, enzyme reactions, and so on.

In response to the above concerns, much effort has been invested on the inclusion of explicit polarization within the molecular mechanics potentials.[7-9] Several methods are available to explicitly model polarization in molecular simulations, such as the Drude oscillator,[10, 11] fluctuating charges,[12] and induced dipoles.[6, 13, 14] The use of polarizable point dipoles is a classical approach with a long history in molecular simulation.[15] The original induced dipole model of Applequist places induced point dipoles on atom centers.[16] However, this model suffers from the so-called "polarization catastrophe": when interaction between two mutually interacting induced dipoles with atomic polarizabilities diverge at a finite distance. Thole proposed a solution by applying a damping function to induced dipole – induced dipole interactions.[17] However, a drawback to this model is that it does not prescribe how induced dipoles and permanent charges interact. A great deal of effort has been devoted to developing modern polarizable models, including the fluctuating charge models[18, 19] in the context of OPLS-AA, the fluctuating charge model and the Drude oscillator model[20-23] in the context of CHARMM, and detailed multipole expansions and more complicated MM potentials in the context of Amoeba.[24] In Amber, polarization was implemented with induced dipoles.[25] In Amber ff12pol, the induced dipoles are calculated using Thole models to avoid "polarization catastrophe".[26-29]

Another limitation of widely adopted nonpolarizable models is their use of partial atomic charges in the electrostatic models, which often lack sufficient mathematical flexibility to describe the electrostatic potential around molecules. Williams showed that optimal least-squares fitting of atom-centered partial charges resulted in relative root-mean-square errors of 3-10% over a set of grid points in a shell outside the surface of a series of small polar molecules.[30] These errors were reduced by 2-3 orders of magnitude via the use of higher atomic multipoles.[6] In Amoeba force fields, multipoles are placed on each atom, allowing better capture of electrostatic potential distribution around molecules.[31, 32] Gaussian electrostatic model (GEM) is a force field based on density fitting, which can extend to arbitrary angular momentums (multipoles).[33-35] Of course there are many other proposals to model electronic polarization in the literature.[12, 36-38]

Recently, Elking *et al.* proposed to a polarizable multipole model with Gaussian charge densities.[39] A key benefit of the polarizable Gaussian Model (pGM) is its screening of all short-range electrostatic interactions in a physically consistent manner. This is critical for stable charge-fitting in polarizable force fields when the polarization of 1-2 and 1-3 charges are included and are needed to reproduce molecular anisotropy, as we discussed before.[40] An advantage of pGM is that each atom's multipoles are represented by a single Gaussian function and its derivatives with different amplitudes. Therefore, pGM is a minimalist Gaussian polarizable model. In comparison, the GEM model[33-35] treats nuclear charges explicitly and uses Hermite Gaussian auxiliary basis sets to reproduce atomic electron density, so it has the potential to represent the short-range interactions more faithfully than the pGM model. However, because the computational cost of the nonbonded electrostatic calculation scales as the squared number of functions on each atom, the multiple functions used to represent each atom in GEM can notably increase the simulation cost. The increased number of parameters associated with the functions may also pose additional challenges in parameterization. Another major difference is that GEM,[33-35] like several other efforts, such as X-Pol[41] and Amoeba,[31, 32] uses electronic densities to model molecular polarization and other effects. In comparison, our pGM model follows the Amber tradition and uses the *ab initio* electrostatic potential (ESP) to fit the parameters of atomic partial charges and dipoles.

Most macromolecular simulations with long-range electrostatic interactions are performed using periodic boundary conditions. A rigorous treatment of electrostatic interactions in periodic boundary conditions requires a careful treatment of the associated lattice sums. Thus, the widely used lattice sum methods, like particle mesh Ewald (PME), need to be extended to handle multipolar related summations. Fortunately, efficient implementations of PME of dipoles and higher multipoles are already available in widely used software package such as Amber.[42, 43] This greatly simplifies the integration of pGM with PME for molecular simulations.

In the following sections we first describe the detailed pGM electrostatics scheme with a focus on how to define the atomic Gaussian multipoles and associated analytical algorithms for force computation. This is followed by algorithmic details of interfacing pGM and PME. We then present the validation of the analytical force formulation and accuracy discussion of pGM in PME simulations. Finally, we conclude the manuscript with a brief discussion of the next steps in our development.

## 2. Theory

### *2.1 Gaussian Density Representation of Charge Distribution*

The Gaussian multipole model represents the charge distribution on each atom as a Gaussian-shaped multipole expansion. So, an $n^{th}$ order Gaussian multipole with the radius of $1/\beta$, located at position $\vec{R}$, is[39]

$$\rho^{(n)}(\vec{r};\vec{R}) = \Theta^{(n)} \cdot \nabla_R^{(n)} (\frac{\beta}{\sqrt{\pi}})^3 exp\,(-\beta^2|\vec{r}-\vec{R}|^2). \qquad (1)$$

Here $\Theta^{(n)}$ is the $n^{th}$ rank momentum tensor and $\nabla^{(n)}$ is the $n^{th}$ rank gradient operator (Appendix A.1).

In our current polarizable Gaussian multipole (pGM) model, only monopoles and dipoles are retained, so only the first two terms are needed at each atom as shown below

$$\rho^{(0)}(\vec{r}; \vec{R}) = q(\frac{\beta}{\sqrt{\pi}})^3 \exp(-\beta^2|\vec{r} - \vec{R}|^2)$$

$$\rho^{(1)}(\vec{r}; \vec{R}) = \vec{\mu} \cdot \nabla_R (\frac{\beta}{\sqrt{\pi}})^3 \exp(-\beta^2|\vec{r} - \vec{R}|^2) \tag{2}$$

where the zeroth-order term represents a monopole, and the first-order term represents a dipole.

Once the charge densities are defined as in Eqn (2), the pairwise Coulombic interaction energy expressions needed for the current pGM model are as follows

(1) Monopole-Monopole:

$$q_1 q_2 \frac{erf(\beta_{12}R^{12})}{R^{12}} \tag{3}$$

(2) Monopole-Dipole:

$$q_1 \vec{\mu}_2 \cdot \nabla_2 \frac{erf(\beta_{12}R^{12})}{R^{12}} \tag{4}$$

(3) Dipole-Dipole:

$$(\vec{\mu}_1 \cdot \nabla_1)(\vec{\mu}_2 \cdot \nabla_2) \frac{erf(\beta_{12}R^{12})}{R^{12}} \tag{5}$$

where *erf*() is the error function and

$$\beta_{12} = \frac{\beta_1 \beta_2}{\sqrt{\beta_1^2 + \beta_2^2}} \text{ and } R^{12} = |\vec{R_1} - \vec{R_2}| \tag{6}$$

Finally it is often convenient to introduce the dipole-dipole interaction tensor $\vec{\vec{T}}_{12} = \nabla_1 \nabla_2 \frac{erf(\beta_{12}R^{12})}{R^{12}}$ so that Eqn (5) can be simplified as $\vec{\mu}_1 \cdot \vec{\vec{T}}_{12} \cdot \vec{\mu}_2$. Here it is worth pointing out an important convention used throughout this manuscript. All gradient operators paired with a dipole only operate on coordinates. For example, the gradient operator in $\vec{\mu}_1 \cdot \nabla_1$ only operates on the atomic coordinates that follow. On the other hand, all other gradient operators that are not paired with a dipole are used in the normal sense. This convention is adopted throughout this manuscript.

It can be shown that an effective potential and corresponding effective field at atomic center $\vec{R_1}$ can be defined as

$$\phi^{effective} = (q_2 + \vec{\mu}_2 \cdot \nabla_2) \frac{erf(\beta_{12}R^{12})}{R^{12}}$$

$$E^{effective} = -(q_2 + \vec{\mu}_2 \cdot \nabla_2) \nabla_1 \frac{erf(\beta_{12}R^{12})}{R^{12}} \tag{7}$$

due to a charge distribution at atomic center $\vec{R_2}$, so that the pairwise Coulomb energies in Eqn. (3) through (5) can be reproduced when an effective point charge of $q_1$ and an effective point dipole $\vec{\mu}_1$ are placed at atomic center $\vec{R_1}$. The use of effective potential and field simplifies the derivation of pairwise Coulombic force calculations as to be shown below. Note that these are different from the real Coulombic potential and field due to Gaussian charges and dipoles. For example, the real potential at any location $\vec{R_1}$ due to atom 2 at $\vec{R_2}$ is

$$\phi^{real} = (q_2 + \vec{\mu}_2 \cdot \nabla_2) \frac{erf\ (\beta_2 R^{12})}{R^{12}} \tag{8}$$

*2.2 Gaussian Multipoles in pGM*

In our current pGM model, interactions are modeled with both permanent and induced atomic multipoles at atomic centers, both of which are truncated at the dipole level. The framework can be easily extended to higher-order multipoles, if needed in future developments.

*Permanent Multipoles* Permanent multipoles are the first part of the pGM model and are defined with respect to a local frame overlapped with an atom's covalent bonds. This choice is based on the fact that atomic moments result from atomic covalent bonding interactions. This is also because covalent bonding interactions are along the stiffest degrees of freedom of a molecule. Thus our design follows the logic that induced moments are meant to be responsible for changes in molecular moments due to changes in soft degrees of freedom in molecular simulations. Of course, the partition between permanent and induced moments is somewhat artificial in a moment fitting procedure. Therefore we refer permanent multipoles in our pGM model as covalent multipoles in the following discussion.

The zeroth-order covalent multipoles, i.e. covalent monopoles, are simply the atomic partial charges as in other polarizable or nonpolarizable force fields. The first-order multipoles, i.e. covalent dipoles, are expressed in linear combinations of certain basis vectors. We define the basis vectors to be along the bonding directions, or more precisely, covalent interaction directions. Thus, there may be more covalent interactions than the number of bonds needed to fully define all covalent dipoles on an atom. For example, hydrogen atoms in water are with covalent dipole moments not 100% along the H-O bonds, so virtual H-H bonds may be needed to define covalent dipoles more accurately. An illustration of basis vectors is shown in Figure 1 for O and H atoms of water, and we refer these as the covalent basis vectors (CBVs). The local frame formed by CBVs on an atom is termed its CBV frame. Another representative case is the alpha carbon atom in proteins, which has four bonds, so there can be four basis vectors in its CBV frame to define the covalent dipoles.

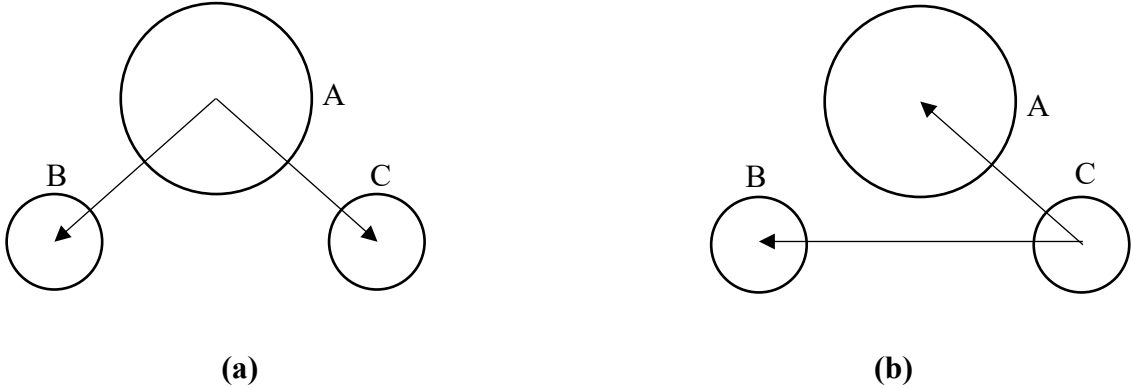

**Figure 1.** Definition of covalent basis vectors for atoms in the water molecule. (a) For covalent dipoles centered at A (oxygen), the two basis vectors are $\vec{e}^{BA}$ and $\vec{e}^{CA}$, which are defined as unit vectors along its two O-H bonds. (b) For covalent dipoles centered at C (hydrogen), the two basis vectors are $\vec{e}^{AC}$, the unit vector along the H-O bond, and $\vec{e}^{BC}$, the unit vector along the H-H virtual bond. The covalent dipoles centered on the other hydrogen atom B can be defined similarly.

The CBV frame is also chosen for the sake of simplifying force calculations because the basis vectors are directly dependent on the positions of atoms. For example, the gradient of a covalent dipole vector used extensively in force calculations can be obtained easily within the CBV frame as

$$\nabla(\vec{u}) = \nabla\left(\sum_i u_i \frac{\vec{R}_i}{|\vec{R}_i|}\right) = \sum_i u_i \left(\frac{\vec{\vec{I}}}{|\vec{R}_i|} - \frac{\vec{R}_i \vec{R}_i}{|\vec{R}_i|^3}\right), \qquad (9)$$

where $\vec{u}$ is the permanent dipole of an atom, $\vec{R}_i$ is the vector pointing from the atom to its $i^{th}$ bonded atom (including virtually), $\vec{\vec{I}}$ is the identity tensor, and the summation is over all covalent interactions of the atom.

Even if quadrupoles are not used in the current pGM model, it is instructive to outline how they are defined in the CBV frame. Given the covalent basis vectors defined above, covalent basis tensors are constructed as dyadic tensors, with each of which formed as a dyadic product of two covalent basis vectors. For example in the case of oxygen atom with two covalent basis vectors ($\vec{e}^{BA}, \vec{e}^{CA}$) in Figure 1a, there are up to four dyadic tensors ($\vec{e}^{BA}\vec{e}^{BA}, \vec{e}^{CA}\vec{e}^{CA}, \vec{e}^{BA}\vec{e}^{CA}, \vec{e}^{CA}\vec{e}^{BA}$) available to define its quadrupole.

*Induced Multipoles* Induced multipoles are the second part of the pGM model. Only first-order terms, i.e. induced dipoles, are used. The pGM polarization scheme can naturally avoid the well-known polarization catastrophe in point polarizable models without employing any artificial screening factors,[17] because distributed dipole densities instead of point dipoles are induced at atomic centers.[44]

In the current pGM model, the linear polarization relation is retained as follows

$$\vec{p}_i = \alpha_i \vec{E}_i^{effective} = \alpha_i(\vec{E}_i^{covalent,effective} - \sum_{j \neq i} \vec{\vec{T}}_{ij} \cdot \vec{p}_j)$$

$$\vec{E}_i^{covalent,effective} = -\sum_{j \neq i}(q_j + \vec{\mu}_j \cdot \nabla_j)\nabla_i \frac{\text{erf}(\beta_{ij}R^{ij})}{R^{ij}}$$

$$\vec{\vec{T}}_{ij} = \nabla_i \nabla_j \frac{\text{erf}(\beta_{ij}R^{ij})}{R^{ij}} \tag{10}$$

where $\vec{p}_i$ is the induced dipole and $\alpha_i$ is the polarizability coefficient of atom *i*, $\vec{E}_i^{effective}$ is the total effective electric field at atom *i*, which contains two parts, (1) the effective field of covalent multipoles, $\vec{E}_i^{covalent,effective}$ (Eqn (7)), and (2) that of induced dipoles, $-\sum_{j \neq i} \vec{\vec{T}}_{ij} \cdot \vec{p}_j$. Note that we have used the effective electric field instead of the real electric field to define the induced dipoles in the pGM model. One reason is to ensure the symmetry of $\vec{\vec{T}}_{ij}$, which greatly reduces the complexity of force calculation later.[44] To simplify the following discussion, we drop the *effective* superscript as we plan to use effective electric fields in all subsequent discussions of energy and force calculations in the pGM model.

Another issue worth pointing out about induced dipoles is their self energies. The linear polarization itself implies a self-energy term of the form

$$U = \frac{1}{2}\frac{\vec{p}^2}{\alpha}. \tag{11}$$

The derivation can be found in many publications.[45] However, the pGM model, due to its use of Gaussian distributions of multipoles, posts extra difficulty. For example, a Gaussian charge distribution itself has self energy, or assembly energy, of the form as derived in Appendix A.2 as

$$U = \frac{q^2\beta}{\sqrt{2\pi}} + \frac{\beta^3(\vec{\mu}+\vec{p})^2}{3\sqrt{2\pi}} \tag{12}$$

Clearly, the self energy is different from Eqn. (11) and it does not lead to a linear polarization behavior. In fact, it is difficult to assess the physical meaning for the nonlinear assembly energy, just like it is hard to discuss the physical meaning of the infinitely large assembly/self energy for a point charge. Thus, for the current model development, we do not consider self energies beyond Eqn. (11).

*2.3 Total Electrostatic Energy and Forces in pGM model*

From the introduction of the pGM model in previous sections, it is clear that the electrostatic potential energy of the system can be divided into two parts:

(1) Covalent Dipole-Covalent Dipole Interaction Energy:

$$U_{covalent-covalent} = \frac{1}{2}\sum_i^N \sum_{j\neq i}^N (q_i + \vec{\mu}_i \cdot \nabla_i)(q_j + \vec{\mu}_j \cdot \nabla_j)\frac{\text{erf}(\beta_{ij}R^{ij})}{R^{ij}} \qquad (13)$$

(2) Induced Energy:

$$U_{induced} = U_{induced-covalent} + U_{induced-induced} + U_{self}$$

$$= -\sum_i^N \vec{p}_i \cdot \vec{E}_i^{covalent} + \frac{1}{2}\sum_i^N \sum_{j\neq i}^N (\vec{p}_i \cdot \vec{\vec{T}}_{ij} \cdot \vec{p}_j) + \frac{1}{2}\sum_i^N \vec{p}_i \cdot (\vec{E}_i^{covalent} - \sum_{j\neq i}^N \vec{\vec{T}}_{ij} \cdot \vec{p}_j)$$

$$= -\frac{1}{2}\sum_i^N \vec{p}_i \cdot \vec{E}_i^{covalent} \qquad (14)$$

where *N* denotes the number of atoms in the system. Thus, atomic electrostatic forces can be derived as negative gradients of the above two potential energy terms, respectively.

When computing gradients for the covalent-dipole interaction energy, it is very important to know which quantities are the variables of the virtual displacement of atom *i*. There are two types of variables: (1) pairwise distances between atom *i* and all other atoms, and (2) covalent dipoles on atom *i* and covalent dipoles on atoms covalently interacting with atom *i*. Given this classification of variables, we can group the terms in Eqn. (13) into four different parts and discuss their gradients with respect to $\vec{R}_i$, separately.

The detailed derivations are presented in Appendix A.3, and the final force expression for the covalent-dipole interaction energy is

$$\vec{F}_{covalent-covalent}^i = -\sum_j^N \nabla_i(\vec{u}_j) \cdot \sum_{k\neq j}^N \nabla_j(q_k + \vec{u}_k \cdot \nabla_k)\frac{\text{erf}(\beta_{jk}R^{jk})}{R^{jk}}$$

$$-(q_i + \vec{u}_i \cdot \nabla_i)\sum_{j\neq i}^N (q_j + \vec{u}_j \cdot \nabla_j)\nabla_i\frac{\text{erf}(\beta_{ij}R^{ij})}{R^{ij}} \qquad (15)$$

where $N$ denotes the number of atoms in the system. Briefly, the first term is from the derivatives over the covalent dipoles, and the second term is from derivatives over the pairwise distances.

The derivation of forces for the induced energy in Eqn (14) is not that straightforward. We first need to express the induced dipole, $\vec{p}_i$, in terms of fields from covalent dipoles only, not as their definition in Eqn (10). This is because $\vec{p}_i$ appears on both sides of Eqn (10), i.e. induced dipoles mutually influence each other, so it is difficult to take their derivatives. Instead we proceed by expressing $\vec{p}_i$ as shown in Appendix A.3 as

$$p_i^s = A^{-1}{}_{ij}^{st} E_j^{covalent,t} \tag{16}$$

where $A^{-1}{}_{ij}^{st}$ is an $3N \times 3N$ matrix which we do not know the expression of, and $s\,t$ and $i\,j$ are coordinate component indices and atom indices, respectively. Next Eqn (14) can be rewritten as

$$U = -\frac{1}{2} A^{-1}{}_{jk}^{st} E_j^{covalent,s} E_k^{covalent,t} \tag{17}$$

where the Einstein's index notation is employed for $j\,k\,s\,t$, so that a repeated index implies a summation over all possible values of the index, i.e. Eqn (17) is a quadruple summation. Even if we do not know the expression of matrix $A^{-1}$, we can still obtain its gradient with respect to $\vec{R}_i$, or the virtual displacement of atom $i$

$$\frac{\partial A^{-1}{}_{ij}^{st}}{\partial x_k^w} = -A^{-1}{}_{ii'}^{ss'} \frac{\partial T_{i'j'}^{s't'}}{\partial x_k^w} A^{-1}{}_{j'j}^{t't} \tag{18}$$

where all primed indices follow the Einstein's index notation, and $T$ is the dipole-dipole interaction tensor.

Given the above preparations, the induced part of the force can be obtained as

$$\vec{F}_{induced}^i = -\vec{p}_i \cdot \nabla_i \sum_{j \neq i}^{N} (\vec{p}_j \cdot \nabla_j) \nabla_i \frac{\text{erf}(\beta_{ij} R^{ij})}{R^{ij}} + \sum_{j}^{N} \nabla_i (\vec{E}_j^{covalent}) \cdot \vec{p}_j \tag{19}$$

where $N$ denotes the number of atoms in the system. Details are presented in Appendix A.3. Briefly, the first term is obtained from the derivative of matrix $A^{-1}$ and the second term is calculated as derivative of the covalent field $\vec{E}_j^{covalent}$ as follows

$$\nabla_i (\vec{E}_j^{covalent}) =$$

$$\begin{cases} -\sum_{k \neq j}^{n} \nabla_i(\vec{u}_k) \cdot \nabla_k \nabla_j \frac{\text{erf}(\beta_{jk} R^{jk})}{\beta_{jk} R^{jk}} - (q_i + \vec{u}_i \cdot \nabla_i) \nabla_i \nabla_j \frac{\text{erf}(\beta_{ij} R^{ij})}{\beta_{ij} R^{ij}} & \text{if } j \neq i \\ -\sum_{k \neq i}^{n} \nabla_i(\vec{u}_k) \cdot \nabla_k \nabla_i \frac{\text{erf}(\beta_{ik} R^{ik})}{\beta_{ik} R^{ik}} - \sum_{k \neq i}^{N} (q_k + \vec{u}_k \cdot \nabla_k) \nabla_i \nabla_i \frac{\text{erf}(\beta_{ik} R^{ik})}{\beta_{ik} R^{ik}} & \text{if } j = i \end{cases} \tag{20}$$

where *n* denotes all atoms that are covalently (including virtually) bonded with atom *i*, and atom *i* itself.

In practice, force calculations must be combined with an Ewald summation or particle mesh Ewald (PME) technique to handle long-range under periodic boundary condition. This is to be discussed in detail in section 2.5.

## 2.4 Ewald Summation and PME in pGM

The Ewald summation was introduced to compute the electrostatic energy of an infinite lattice under periodic boundary conditions.[46] The basic idea is to put a mask Gaussian charge distribution on the real charge on each atom. Then a direct-space pairwise summation is conducted to compute the electric field due to real charges masked by the Gaussian charges. This step can be executed with a reasonably short cutoff distance due to the very fast decay after applying the mask Gaussian charges. Next the field generated by the mask Gaussians can be computed efficiently by a reciprocal-space summation to bring back the original electric field due to the real charges. Finally, a correction step is used to remove interactions not needed in the original electrostatic model. Similar to its use in point charge/dipole models, a mask Gaussian distribution is also used on each moment of each atom in the pGM model.

$$\rho_i^{mask}(\vec{r}; \vec{R}_i) = q_i \left(\frac{\beta_0^2}{\pi}\right)^{\frac{3}{2}} \exp\left(-\beta_0^2 |\vec{r} - \vec{R}_i|^2\right)$$

$$+ (\vec{\mu}_i + \vec{p}_i) \cdot \nabla_{R_i} \left(\frac{\beta_0^2}{\pi}\right)^{\frac{3}{2}} \exp\left(-\beta_0^2 |\vec{r} - \vec{R}_i|^2\right) \tag{21}$$

where $\beta_0$ is an adjustable parameter usually in the range of about $\frac{1}{5} \sim \frac{1}{2}$ Å$^{-1}$, universal for all atoms.

*Direct Summation* Given the mask distribution is also a Gaussian function, it is straightforward to compute electrostatic potential, field, and the gradient of field of a masked pGM charge distribution, as follows

$$\phi_i = \sum_{j \neq i}^{N} [q_j + (\vec{\mu}_j + \vec{p}_j) \cdot \nabla_j] \frac{\text{erf}(\beta_{ij} R^{ij}) - \text{erf}(\beta_0 R^{ij})}{R_{ij}} \tag{22}$$

$$\vec{E}_i = -\sum_{j \neq i}^{N} [q_j + (\vec{\mu}_j + \vec{p}_j) \cdot \nabla_j] \nabla_i \frac{\text{erf}(\beta_{ij} R^{ij}) - \text{erf}(\beta_0 R^{ij})}{R^{ij}} \tag{23}$$

$$\vec{E}_i = -\sum_{j \neq i}^{N} [\, q_j + (\vec{\mu}_j + \vec{p}_j) \cdot \nabla_j\,] \nabla_i \nabla_i \frac{\text{erf}(\beta_{ij}R^{ij}) - \text{erf}(\beta_0 R^{ij})}{R_{ij}} \qquad (24)$$

For the current pGM model, no higher-order field is needed. Here, $N$ represents all the atoms including those in the periodic boxes, but their influence would decay to zero very quickly due to the masking effect.

Another point worth pointing out is that the real field of the mask Gaussian multipoles is used, i.e. $\beta_0$ is used instead of $\beta_{i0} = \frac{\beta_i \beta_0}{\sqrt{\beta_i^2 + \beta_0^2}}$ in the above expressions. The mixed use is not an issue because mask multipoles do not really exist, their role is just a mathematical treatment in the Ewald summation as long as the effect is exactly cancelled out in the later step.

*Reciprocal Summation* The reciprocal summation of the pGM model follows the same procedure as a traditional point polarizable model.[43] Thus the electrostatic potential, field, and gradient of field can be shown as

$$\phi_i = \frac{1}{\pi V} \sum_{\vec{m} \neq 0} \frac{\exp\left(-\frac{\pi^2 \vec{m}^2}{\beta_0^2}\right)}{\vec{m}^2} \exp(-2\pi i \vec{m} \cdot \vec{R}_i) S(\vec{m}) \qquad (25)$$

$$\vec{E}_i = \frac{2i}{V} \sum_{\vec{m} \neq 0} \vec{m} \frac{\exp\left(-\frac{\pi^2 \vec{m}^2}{\beta_0^2}\right)}{\vec{m}^2} \exp(-2\pi i \vec{m} \cdot \vec{R}_i) S(\vec{m}) \qquad (26)$$

$$\vec{\vec{E}}_i = -\frac{4\pi}{V} \sum_{\vec{m} \neq 0} \vec{m}\vec{m} \frac{\exp\left(-\frac{\pi^2 \vec{m}^2}{\beta_0^2}\right)}{\vec{m}^2} \exp(-2\pi i \vec{m} \cdot \vec{R}_i) S(\vec{m}) \qquad (27)$$

where the $i$'s that are not subscripts but the imaginary unit, V is the volume of the unit cell, and $\vec{m}$ is the reciprocal space vector. $S(\vec{m})$ is the structure factor

$$S(\vec{m}) = \sum_{j=1}^{N} \tilde{L}_j(\vec{m}) \exp(2\pi i \vec{m} \cdot \vec{R}_j)$$
$$\tilde{L}_j(\vec{m}) = q_j + 2\pi i (\vec{\mu}_j + \vec{p}_j) \cdot \vec{m} \qquad (28)$$

Here $N$ is the number of atoms in the primary simulation box only.

*Correction* The PME correction term is used to handle various specific situations in a force field. For example, most force fields have masked bonded (1-2 and 1-3) atom pairs, which result in no electrostatic interactions between these pairwise atoms. Thus, the interactions among these masked pairs must be removed. However, in the current pGM model we do not have any masked pairs, so there is no need for such correction.

Another correction that needs paying attention to is the self-interaction correction. This is the only correction in the pGM model. The self-potential, self-field, gradient of self-field can be shown as[42, 43]

$$\phi_i = \frac{2q_i\beta_0}{\sqrt{\pi}} \tag{29}$$

$$\vec{E}_i = -\frac{4(\vec{\mu}_i+\vec{p}_i)\beta_0^3}{3\sqrt{\pi}} \tag{30}$$

$$\vec{\vec{E}}_i = \frac{4q_i\beta_0^3}{3\sqrt{\pi}}\vec{\vec{I}} \tag{31}$$

These terms need to be properly subtracted to obtain correct potential, field, and field gradient, respectively

In summary, the similarity between the Ewald summation in pGM model and that in the polarizable point charge/dipole model shows that the PME MD engine for the point charge/dipole model can be easily transplanted over for pGM applications with little revision. There are excellent literature discussing the details of PME for polarizable point dipole models, and can be safely omitted in this work.[42, 43]

## *2.5 Computing Forces with Ewald Summation and PME*

As pointed out at the end of Section 2.3, analytical force expressions, Eqn (15) and (19), cannot be used directly in typical MD simulations since all summations are over infinite numbers of atoms with periodic boundary conditions. They must be combined with an Ewald summation or a PME technique to facilitate solvated-phase simulations. To bypass the infinite summations, the force expressions are reformulated in terms of fields and its derivatives, which are also the quantities that an Ewald or PME procedure would return. The key to express Eqn (15) and (19) with fields and gradients of fields is to consider the following quantities together

$$\vec{E}_i^{covalent} = -\sum_{j\neq i}^{N}(q_j + \vec{\mu}_j \cdot \nabla_j)\nabla_i \frac{\text{erf}(\beta_{ij}R^{ij})}{R^{ij}}$$

$$\vec{\vec{E}}_i^{covalent} = -\sum_{j\neq i}^{N}(q_j + \vec{\mu}_j \cdot \nabla_j)\nabla_i\nabla_i \frac{\text{erf}(\beta_{ij}R^{ij})}{R^{ij}}$$

$$\vec{E}_i^{induced} = -\sum_{j\neq i}^{N}\vec{p}_j \cdot \nabla_j\nabla_i \frac{\text{erf}(\beta_{ij}R^{ij})}{R_{ij}}$$

$$\vec{\vec{E}}_i^{induced} = -\sum_{j\neq i}^{N}\vec{p}_j \cdot \nabla_j\nabla_i\nabla_i \frac{\text{erf}(\beta_{ij}R^{ij})}{R^{ij}} \tag{32}$$

where *N* denotes all atoms in the system, including those of the periodic boxes. Thus, these are all infinite summations.

A key step is in the computation of $\vec{F}^i_{induced}$, where term $\sum_j^N \nabla_i(\vec{E}^{covalent}_j) \cdot \vec{p}_j$ also has to be reformulated accordingly. Given that Eqn (20) for $\nabla_i(\vec{E}^{covalent}_j)$ lists two separate terms for $j = i$ and $j \neq i$, we can rewrite $\sum_j^N \nabla_i(\vec{E}^{covalent}_j) \cdot \vec{p}_j$ as follows

$$\sum_j^N \nabla_i(\vec{E}^{covalent}_j) \cdot \vec{p}_j$$
$$= \sum_{j\neq i}^N \sum_{k\neq j}^n (-\nabla_i(\vec{u}_k)) \cdot \nabla_k(\vec{p}_j \cdot \nabla_j) \frac{\text{erf}(\beta_{jk}R^{jk})}{\beta_{jk}R^{jk}} - \sum_{j\neq i}^N (q_i + \vec{\mu}_i \cdot \nabla_i) \nabla_i(\vec{p}_j \cdot \nabla_j) \frac{\text{erf}(\beta_{ij}R^{ij})}{\beta_{ij}R^{ij}}$$
$$- \sum_{k\neq i}^n \nabla_i(\vec{u}_k) \cdot \nabla_k(\vec{p}_i \cdot \nabla_i) \frac{\text{erf}(\beta_{ik}R^{ik})}{\beta_{ik}R^{ik}} - \sum_{k\neq i}^N (q_k + \vec{\mu}_k \cdot \nabla_k) \nabla_i(\vec{p}_i \cdot \nabla_i) \frac{\text{erf}(\beta_{ik}R^{ik})}{\beta_{ik}R^{ik}} \quad (33)$$

where both $j = i$ and $j \neq i$ terms in Eqn (20) are needed due to the outermost summation over *j*. Combining the first and third terms of Eqn (33) gives

$$= \sum_j^N \sum_{k\neq j}^n (-\nabla_i(\vec{u}_k)) \cdot \nabla_k(\vec{p}_j \cdot \nabla_j) \frac{\text{erf}(\beta_{jk}R^{jk})}{\beta_{jk}R^{jk}} - \sum_{j\neq i}^N (q_i + \vec{\mu}_i \cdot \nabla_i) \nabla_i(\vec{p}_j \cdot \nabla_j) \frac{\text{erf}(\beta_{ij}R^{ij})}{\beta_{ij}R^{ij}}$$
$$- \sum_{k\neq i}^N (q_k + \vec{\mu}_k \cdot \nabla_k) \nabla_i(\vec{p}_i \cdot \nabla_i) \frac{\text{erf}(\beta_{ik}R^{ik})}{\beta_{ik}R^{ik}} \quad (34)$$

Exchanging the summation order for the first term leads to

$$= \sum_k^n \sum_{j\neq k}^N (-\nabla_i(\vec{u}_k)) \cdot \nabla_k(\vec{p}_j \cdot \nabla_j) \frac{\text{erf}(\beta_{jk}R^{jk})}{\beta_{jk}R^{jk}} - \sum_{j\neq i}^N (q_i + \vec{\mu}_i \cdot \nabla_i) \nabla_i(\vec{p}_j \cdot \nabla_j) \frac{\text{erf}(\beta_{ij}R^{ij})}{\beta_{ij}R^{ij}} -$$
$$\sum_{k\neq i}^N (q_k + \vec{\mu}_k \cdot \nabla_k) \nabla_i(\vec{p}_i \cdot \nabla_i) \frac{\text{erf}(\beta_{ik}R^{ik})}{\beta_{ik}R^{ik}} \quad (35)$$

Substitution of the expressions of electric fields and derivatives in Eqn (32) gives

$$\sum_j^N \nabla_i(\vec{E}^{covalent}_j) \cdot \vec{p}_j =$$

$$\sum_k^n \nabla_i(\vec{u}_k) \cdot \vec{E}^{induced}_k + q_i \vec{E}^{induced}_i + \vec{u}_i \cdot \vec{E}^{induced}_i + \vec{p}_i \cdot \vec{E}^{covalent}_i \quad (36)$$

Here, *n* means those atoms that covalently interact with atom *i*.

Given the above preparations, Eqn (15) and (19) can finally be expressed as follows after substitution of Eqns (32) and (36)

$$\vec{F}^i_{covalent-covalent} = \sum_j^n \nabla_i(\vec{u}_j) \cdot \vec{E}^{covalent}_j + q_i \vec{E}^{covalent}_i + \vec{u}_i \cdot \vec{E}^{covalent}_i \quad (37)$$
$$\vec{F}^i_{induced} = \sum_j^n \nabla_i(\vec{u}_j) \cdot \vec{E}^{induced}_j + q_i \vec{E}^{induced}_i$$

$$+\vec{p}_i \cdot \vec{E}_i^{induced} + \vec{u}_i \cdot \vec{E}_i^{induced} + \vec{p}_i \cdot \vec{E}_i^{covalent} \tag{38}$$

Adding these two terms together, the final force expression is,

$$\vec{F}^i = \sum_j^n \nabla_i \left(\vec{u}_j\right) \cdot \vec{E}_j + q_i \vec{E}_i + (\vec{u}_i + \vec{p}_i) \cdot \vec{\vec{E}}_i \tag{39}$$

Eqn (39) shows that a key step in this algorithm is to accumulate atomic electric potential and its first and second derivatives from various components, including both reciprocal and direct summations.

## 3. Results and Discussion

### 3.1 Validation of Analytical Electrostatic Force Expression

To validate the pGM force expressions, e.g. Eqns (15)/(19) above, we constructed a small toy system of two water molecules in free space. The detailed water pGM parameters are listed in Table 1. These parameters were derived with an iterative RESP procedure for the pGM model with quantum mechanical ESP data from a B3LYP/aug-cc-pvtz calculation of the water dimer.

| Tested atoms | water-1 O | water-1 H1 | water-1 H2 | water-2 O | water-2 H1 | water-2 H2 |
|---|---|---|---|---|---|---|
| Charge ($e$) | -1.797045 | 0.898523 | 0.898523 | -1.797045 | 0.898523 | 0.898523 |
| Dipole moment ($e \cdot Å$) | (1.317332, -0.120867, 1.711988) | (0.262849, -0.484213, 0.323876) | (0.276182, 0.436999, 0.376728) | (0.927376, 0.134058, -1.967337) | (-0.236591, 0.041394, -0.650790) | (0.590269, 0.013966, -0.164059) |
| Polarizability ($Å^3$) | 1.448980 | 0.427350 | 0.427350 | 1.448980 | 0.427350 | 0.427350 |
| Gaussian radius (Å) | 0.8066249 | 0.7147597 | 0.7147597 | 0.8066249 | 0.7147597 | 0.7147597 |
| Coordinates (Å) | (-1.387669, -0.006775, 0.110728) | (-1.734110, 0.790147, -0.310036) | (-1.756164, -0.740122, -0.397708) | (1.514536, 0.007522, -0.121554) | (1.919419, -0.047517, 0.751251) | (0.555921, -0.008485, 0.043101) |

Table 1. Two water molecules in free space. The permanent dipole moments are expressed in the lab frame. The moments in the CBV frame can be obtained once the CBV's are defined according to Figure 1.

Two methods were used to calculate the atomic forces. The first method is to use the force expressions to calculate forces analytically. The second method is to calculate forces numerically via the finite-difference method, based on the fact that each force is the negative gradient of potential energy. Here the potential energy was computed with Eqn (13) and (14). The finite-difference coordinate displacement was set to be 0.001 Å. The two sets of atomic forces are listed in Table 2. It is clear that the differences between the two sets of atomic forces appear only on the fourth digit after the decimal point, which is consistent with the accuracy of the finite-difference displacement used (0.001 Å).

| Tested atoms | water-1 O | water-1 H1 | water-1 H2 | water-2 O | water-2 H1 | water-2 H2 |
|---|---|---|---|---|---|---|
| Analytical forces | (0.094496, -0.008624, 0.1254230) | (-0.048543, -0.237284, -0.076339) | (-0.041431, 0.245927, -0.048632) | (0.047721, 0.009654, -0.143042) | (-0.245399, 0.000915, -0.033889) | (0.193156, -0.010589, 0.176479) |
| Finite-difference forces | (0.094410, -0.008797, 0.125186) | (-0.048657, -0.237474, -0.076493) | (-0.041549, 0.245762, -0.048810) | (0.047509, 0.009778, -0.143437) | (-0.245485, 0.000871, -0.034221) | (0.192871, -0.010637, 0.176359) |
| Deviations | (0.000085, 0.000173, 0.000236) | (0.000114, 0.000190, 0.000154) | (0.000118, 0.000164, 0.000178) | (0.000211, -0.000124, 0.000395) | (0.000086, 0.000043, 0.000332) | (0.000284, 0.000048, 0.000119) |

Table 2. Atomic forces ($e^2/Å^2$) computed via the analytical expression and the finite difference procedure, and their differences.

There is also an indirect way to confirm the correctness of the force expression, which is to utilize the fact that the total force of the system should always be zero in any direction. If we add up all atomic forces, the system net force (in $e^2/Å^2$) is $-2 \times 10^{-17}$, $-1 \times 10^{-6}$ and 0, for *x*, *y* and *z* directions, respectively. The overall error here is consistent with the induction tolerance used in the testing, $1 \times 10^{-6}$.

**3.2 Accuracy of pGM Electrostatic Energy and Forces in PME**

To achieve aqueous-phase simulations, an Ewald summation or PME technique is essential for any electrostatic model. Although there are various publications discussing the accuracy of PME [42, 43, 47, 48], we have to acknowledge the fact that the pGM model has a higher accuracy requirement than classical point-charge force fields due to the presence of dipoles. In general, higher moments would require higher PME accuracy. This can be appreciated from the perspective of two considerations. Firstly, the potentials of dipoles scale with distance more nonlinearly ($1/r^3$) than those of charges ($1/r$). Secondly, the second derivatives of potential are needed to compute forces on dipoles (Eqn (39)), whereas only first derivatives, e.g. electric fields, are needed to compute forces on charges. Due to these differences, we have to carefully examine the accuracy requirement of PME methods used in our model.

To test the accuracy, we only look at the most difficult pairwise interactions, so that the errors reported below are the maximum errors in the tested water system. For the reciprocal part, we focus on the electrostatic field between the bonded O atom and H atoms, whose interactions are the most nonlinear and thus the most difficult in PME. For the direct summation part, we focus on the electrostatic field between two H atoms. Because they have the smallest Gaussian radii, their interactions converge slowest in the direct summation. Thus, to guarantee a given accuracy level for forces in the pGM model, we need to consider both PME components.

In the following analysis we set the grid spacing to 1 Å for PME as in most biomolecular simulations and varied other parameters to see how accuracy changes in both the reciprocal and the direct summation components. Also, because our model contains both charges and dipoles, we analyzed their field separately to assess the impact of different setups on their accuracy of electric field. The test results are shown below in Table 3, 4 and 5.

| Interpolation order | | 5 | 6 | 7 | 8 | 9 |
|---|---|---|---|---|---|---|
| Ewald coefficient $\beta_0 = 0.3\text{Å}^{-1}$ | potential | $8.2 \times 10^{-5}$ | $2.8 \times 10^{-5}$ | $1.5 \times 10^{-7}$ | $3.4 \times 10^{-6}$ | $2.7 \times 10^{-6}$ |
| | first derivative | $3.0 \times 10^{-4}$ | $6.4 \times 10^{-5}$ | $9.2 \times 10^{-6}$ | $7.6 \times 10^{-6}$ | $3.9 \times 10^{-6}$ |
| | second derivative | $1.3 \times 10^{-3}$ | $3.9 \times 10^{-4}$ | $4.6 \times 10^{-5}$ | $1.6 \times 10^{-5}$ | $5.7 \times 10^{-6}$ |
| Ewald coefficient $\beta_0 = 0.4\text{Å}^{-1}$ | potential | $1.2 \times 10^{-4}$ | $4.0 \times 10^{-5}$ | $4.3 \times 10^{-6}$ | $3.5 \times 10^{-8}$ | $5.4 \times 10^{-6}$ |
| | first derivative | $1.3 \times 10^{-3}$ | $4.6 \times 10^{-4}$ | $1.5 \times 10^{-4}$ | $6.5 \times 10^{-5}$ | $2.5 \times 10^{-5}$ |
| | second derivative | $2.6 \times 10^{-3}$ | $1.6 \times 10^{-3}$ | $4.0 \times 10^{-4}$ | $1.9 \times 10^{-4}$ | $8.1 \times 10^{-5}$ |
| Ewald coefficient $\beta_0 = 0.5\text{Å}^{-1}$ | potential | $2.6 \times 10^{-4}$ | $1.1 \times 10^{-4}$ | $1.7 \times 10^{-4}$ | $6.1 \times 10^{-5}$ | $9.1 \times 10^{-5}$ |
| | first derivative | $5.1 \times 10^{-3}$ | $2.5 \times 10^{-3}$ | $1.2 \times 10^{-3}$ | $7.2 \times 10^{-4}$ | $4.0 \times 10^{-4}$ |
| | second derivative | $8.8 \times 10^{-3}$ | $5.5 \times 10^{-3}$ | $2.4 \times 10^{-3}$ | $1.3 \times 10^{-3}$ | $8.8 \times 10^{-4}$ |

Table 3. Errors of reciprocal potentials and derivatives generated by the dipoles of water-1 H1 atom on water-1 O atom at different PME setups. The analytical values (Eqns (25) – (28)) were calculated with MATLAB. Interpolation order refers to the rank of the Lagrangian interpolation method used in PME. See Ref[48] for details.

| Interpolation order | | 5 | 6 | 7 | 8 | 9 |
|---|---|---|---|---|---|---|
| Ewald coefficient $\beta_0 = 0.3\text{Å}^{-1}$ | potential | $1.4 \times 10^{-6}$ | $3.7 \times 10^{-8}$ | $6.5 \times 10^{-8}$ | $1.4 \times 10^{-8}$ | $5.3 \times 10^{-9}$ |
| | first derivative | $1.8 \times 10^{-4}$ | $3.8 \times 10^{-5}$ | $5.5 \times 10^{-6}$ | $1.1 \times 10^{-6}$ | $3.4 \times 10^{-7}$ |
| | second derivative | $7.7 \times 10^{-4}$ | $2.1 \times 10^{-4}$ | $3.0 \times 10^{-5}$ | $7.8 \times 10^{-6}$ | $1.8 \times 10^{-6}$ |
| Ewald coefficient $\beta_0 = 0.4\text{Å}^{-1}$ | potential | $1.4 \times 10^{-5}$ | $3.7 \times 10^{-6}$ | $1.6 \times 10^{-6}$ | $6.8 \times 10^{-7}$ | $3.2 \times 10^{-7}$ |
| | first derivative | $7.4 \times 10^{-4}$ | $1.1 \times 10^{-4}$ | $6.1 \times 10^{-5}$ | $2.1 \times 10^{-5}$ | $1.0 \times 10^{-5}$ |
| | second derivative | $2.5 \times 10^{-3}$ | $7.7 \times 10^{-4}$ | $2.1 \times 10^{-4}$ | $7.5 \times 10^{-5}$ | $3.0 \times 10^{-5}$ |
| Ewald coefficient $\beta_0 = 0.5\text{Å}^{-1}$ | potential | $8.1 \times 10^{-5}$ | $3.9 \times 10^{-5}$ | $1.9 \times 10^{-5}$ | $1.2 \times 10^{-5}$ | $7.0 \times 10^{-6}$ |
| | first derivative | $2.5 \times 10^{-3}$ | $7.8 \times 10^{-4}$ | $4.5 \times 10^{-4}$ | $2.6 \times 10^{-4}$ | $1.5 \times 10^{-4}$ |
| | second derivative | $6.7 \times 10^{-3}$ | $2.6 \times 10^{-3}$ | $1.1 \times 10^{-3}$ | $6.3 \times 10^{-4}$ | $3.4 \times 10^{-4}$ |

Table 4. Errors of reciprocal potentials and derivatives generated by the charges of water-1 H1 atom on water-1 O atom at different PME setups. The analytical values (Eqns (25) – (28)) were calculated with MATLAB. Interpolation order refers to the rank of the Lagrangian interpolation method used in PME. See Ref[48] for details.

| Cutoff distance (Å) | 7 | 8 | 9 | 10 | 11 |
|---|---|---|---|---|---|
| $\beta_0 = 0.3\text{Å}^{-1}$ | $3.0 \times 10^{-3}$ | $6.9 \times 10^{-4}$ | $1.3 \times 10^{-4}$ | $2.2 \times 10^{-5}$ | $3.1 \times 10^{-6}$ |
| $\beta_0 = 0.4\text{Å}^{-1}$ | $7.5 \times 10^{-5}$ | $6.0 \times 10^{-6}$ | $3.6 \times 10^{-7}$ | $1.5 \times 10^{-8}$ | $4.9 \times 10^{-10}$ |
| $\beta_0 = 0.5\text{Å}^{-1}$ | $7.4 \times 10^{-7}$ | $1.5 \times 10^{-8}$ | $2.0 \times 10^{-10}$ | $1.5 \times 10^{-12}$ | $7.3 \times 10^{-15}$ |

Table 5. Errors of direct summation potentials for Gaussian potentials between two H atoms at different PME setups. These values are the difference between two error functions, $\text{erf}(\beta R_c) - \text{erf}(\beta_0 R_c)$. Here, $R_c$ is the direct summation cutoff distance, and $\beta = 1/(0.7147597 \cdot \sqrt{2})\text{Å}^{-1}$ for the H atom pairs.

It is clear from the above analyses that the pGM model demands a higher accuracy level than classical point-charge models. This is as expected for any electrostatic model with dipoles or higher moments. For the reciprocal part, the field generated by dipoles are more difficult to handle than that of charges in PME. Comparing Tables 3 and 4, we can see that the errors of dipole fields are about twice larger than that of charge fields. Furthermore, Table 3 and 4 show that the second derivatives are the most difficult in PME. Thus, to ensure the accuracy of the reciprocal summation of PME, we need to make sure the second derivatives reach a specified accuracy level. For example, if we use $5 \times 10^{-5}$ as the accuracy threshold, which is a common choice, we have to set Ewald $\beta_0 = 0.3\text{Å}^{-1}$ and the interpolation order 7 or higher in the PME setup (Table 3).

Situations are similar for the direct summation part. To reach a common accuracy threshold of $5 \times 10^{-5}$, if we set $\beta_0 = 0.3\text{Å}^{-1}$ as in the reciprocal part, the direct space cutoff should be set to a relatively longer cutoff distance of 10 Å, as shown in Table 5. Of course, a choice of larger $\beta_0$ (i.e. $0.35\text{Å}^{-1}$) would allow a commonly used cutoff distance of 9 Å. However, this would require a higher interpolation order to achieve the similar level of accuracy.

### 3.3 NVE Simulations of Water Box

Given all the accuracy consideration in Section 3.2, we performed a pure water simulation to test the energy conservation behavior in an NVE run with the PME treatment. The electrostatic parameters were derived from those in Table 1 and transplanted onto the TIP3P water model. We used 512 water molecules in a truncated octahedron box of 27.5 Å. The dimension of the particle mesh grid is $30^3$, so the grid spacing is a bit less than 1 Å. The PME $\beta_0 = 0.35\text{Å}^{-1}$, the real space cutoff was set as 9 Å, and the interpolation order was 8 so that the overall PME error was less than $5 \times 10^{-5}$.

We first tested a range of induction tolerance criteria, ranging from $10^{-3}$ to $10^{-6}$. Our experiments show that $10^{-3}$ and $10^{-4}$ are clearly not sufficient for the induction iteration, leading to decreasing energy throughout the MD simulations. This is consistent with previous findings in the developments of polarizable point dipole models.[42] Specifically, the energy in the NVE run of $10^{-3}$ drifts too fast, so that it is already out of the plotting range at the 100[th] step, the very first data point. The rest of the energy plots over the simulation time are shown below in Figure 2. The initial testing shows that the energy convergence became much better after we tighten the iteration tolerance to $10^{-5}$. Though the total energy still drifts down a little, but much slower than before. Finally, after we tighten it to $10^{-6}$, the total energy is basically conserved. Of course, the total energy fluctuation does exist.

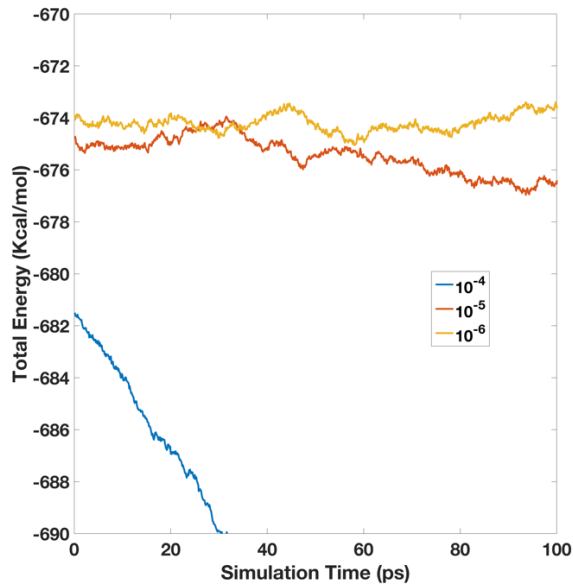

Figure 2. Total energy versus simulation time for the 512-water box simulation. Here, all the simulations are performed with a $5 \times 10^{-5}$ PME accuracy, but with different induction iteration tolerance ($1 \times 10^{-4} \sim 1 \times 10^{-6}$).

Next we also studied the influences of the PME setup on the energy conservation. To compare with the NVE run with the high PME accuracy above. We collected a comparable NVE run with the same induction tolerance of $10^{-6}$, but with a somewhat lower PME setting. The real space cutoff was set as 8 Å and the interpolation order was set as 6 but others remained to be the same, which leads to a lower PME accuracy of $\sim 5 \times 10^{-4}$. The total energy is more positive because there are fewer van der Waals pairs. As shown in Figure 3, the total energy also drifts noticeably, though it becomes more positive over time. In summary, our experiment shows that high enough accuracy in both the PME calculation and the induction iteration is necessary for a polarizable dipole model with permanent dipoles to achieve energy conservation. Furthermore, we expect even higher accuracy is necessary if higher moments, i.e. quadrupoles, are used in future pGM developments as higher derivatives are needed from the PME calculation.

## 4. Conclusion

In this work, we proposed an efficient formulation for the polarizable Gaussian Multipole (pGM) model for biomolecular simulations. Firstly, a local frame based on the covalent basis vectors (CBV) /tensors was used to set up the permanent (covalent) multipoles on all atoms. The CBV frame nicely allows the intrinsic molecular flexibility during simulations and facilitates an efficient formulation of analytical electrostatic forces. Based on the new CBV local frame, we then derived the analytical force expressions for the pGM model. Finally, we outlined how to interface the pGM electrostatics seamlessly with the PME implementation for molecular simulations under the periodic boundary conditions.

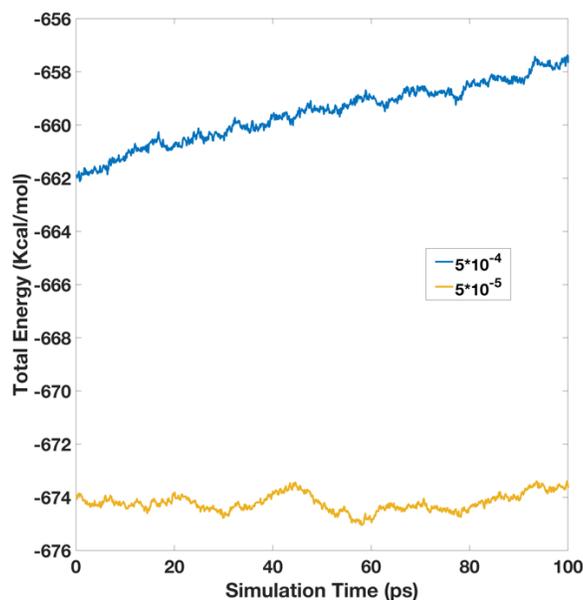

Figure 3. Total energy versus simulation time for the 512-water box simulation. Here, both simulations were performed with a $1 \times 10^{-6}$ induction iteration tolerance, but with different PME accuracy, low for $5 \times 10^{-4}$ and high for $5 \times 10^{-5}$.

To validate the analytical force expression for the pGM model defined on the CBV frame, we studied the accuracy of the analytical atomic forces with a finite-different force analysis for a water dimer. The analysis shows a very good consistency between the analytical and numerical forces, with an error comparable to the finite difference uncertainty. In addition, total analytical and numerical forces of the water dimer are very close to zero with an error consistent with the induction iteration tolerance.

Next, we analyzed the PME setups necessary for accurate pGM energy and force calculations. It was found that the pGM model requires higher accuracy than the classical point-charge models due to the presence of dipoles. This is because the electrostatic field generated by dipoles are much more difficult to interpolate than that of charges in PME, and the error of the dipole field is about twice that of the charge field. In addition, the second derivative of potential is needed, which is the more difficult to compute accurately in PME to ensure accurate pGM forces.

To validate the overall electrostatic framework for the reformulated pGM model, we conducted an NVE simulation for a small water box of 512 water molecules. Our results show that, to achieve energy conservation, it is important to ensure enough accuracy on both PME and induced dipoles. With a $5 \times 10^{-5}$ accuracy on PME and a $1 \times 10^{-6}$ tolerance for induced dipoles, the tested NVE water simulation in the pGM model was shown to conserve energy reasonably well. Future development will be necessary to improve the efficiency of the pGM model in both PME setup and induction iteration to bring out the potential of the pGM model.

## Data Availability

The algorithms developed in this study and the validation data are deposited in the Amber repository and will be made publicly available in the next Amber/AmberTools release on http://ambermd.org/.

## Acknowledgements

This work was supported by NIH GM79383, GM093040, and GM130367.

## Appendices

### A.1 Tensor Format of Boys Serial

In this study, Boys functions up to rank three were used and are listed below as reference. Higher ranked tensors and Boys functions can be found in the literatures.[44, 49]

Boys functions up to rank three are

$$B_0(x) = \frac{\text{erf}(x)}{x}$$

$$B_1(x) = \frac{\text{erf}(x)}{x^3} - \frac{2}{\sqrt{\pi}} e^{-x^2} \frac{1}{x^2}$$

$$B_2(x) = \frac{3\text{erf}(x)}{x^5} - \frac{2}{\sqrt{\pi}} e^{-x^2} \frac{1}{x^4}(3 + 2x^2)$$

$$B_3(x) = \frac{15\text{erf}(x)}{x^7} - \frac{2}{\sqrt{\pi}} e^{-x^2} \frac{1}{x^6}(15 + 10x^2 + 4x^4)$$

The associated tensors are

$$\nabla \frac{\text{erf}(\beta R)}{R} = -\vec{R}\beta^3 B_1(\beta R)$$

$$\nabla\nabla \frac{\text{erf}(\beta R)}{R} = \hat{x}_p \hat{x}_q (R_p R_q \beta^5 B_2(\beta R) - \delta_{pq} \beta^3 B_1(\beta R))$$

$$\nabla\nabla\nabla \frac{\text{erf}(\beta R)}{R} = \hat{x}_p \hat{x}_q \hat{x}_r ((\delta_{pq} R_r + \delta_{pr} R_q + \delta_{rq} R_p)\beta^5 B_2(\beta R) - R_p R_q R_r \beta^7 B_3(\beta R))$$

## A.2 Self-energy derivation

We can derive a general expression of self-energy/assemble energy of Gaussian multipoles in pGM as follows

$$U = \int d\vec{r}\, \frac{1}{2}\rho\varphi = \int d\vec{r}\, \frac{1}{2}(\rho_{monopole} + \rho_{dipole})(\varphi_{monopole} + \varphi_{dipole}) =$$

$$\int d\vec{r}\, \frac{1}{2}\rho_{monopole}\varphi_{monopole} + \int d\vec{r}\, \frac{1}{2}\rho_{dipole}\varphi_{monopole} + \int d\vec{r}\, \frac{1}{2}\rho_{monopole}\varphi_{dipole} +$$

$$\int d\vec{r}\, \frac{1}{2}\rho_{dipole}\varphi_{dipole}$$

The first term, monopole self-energy,

$$\int d\vec{r}\, \frac{1}{2}\rho_{monopole}\varphi_{monopole} = \lim_{R\to 0}\frac{q\cdot q}{2}\frac{erf(\frac{\beta}{\sqrt{2}}R)}{R} = \lim_{R\to 0}\frac{q\cdot q}{2}\frac{\beta}{\sqrt{2}}\frac{erf(\frac{\beta}{\sqrt{2}}R)}{\frac{\beta}{\sqrt{2}}R} = \lim_{x\to 0}\frac{q\cdot q}{2}\frac{\beta}{\sqrt{2}}B_0(x) =$$

$$\frac{q\cdot q}{2}\frac{\beta}{\sqrt{2}}\frac{2}{\sqrt{\pi}} = \frac{q^2\beta}{\sqrt{2\pi}}$$

The second and the third terms, on the other hand, are always zero, because the symmetry of a monopolar and a dipolar distribution is even and odd, respectively, which causes the integral always cancels itself. The last term is the dipole self-energy. In a similar fashion as above for the monopole, it can be shown as

$$\int d\vec{r}\, \frac{1}{2}\rho_{dipole}\varphi_{dipole} = \frac{\beta^3(\vec{\mu}^2+\vec{p}^{\,2})}{3\sqrt{2\pi}}$$

## A.3 Force derivation

We precede in two steps, covalent-covalent interactions and induced interactions as shown in section 2.3. First, we consider interaction energies due to covalent multipoles interacting with covalent multipoles.

The system can be split into two groups of atoms, bonded atoms and nonbonded atoms. Bonded group are those atoms bonded to the atom that are being considered (including itself), and the nonbonded group are the rest. In the bonded group, the atoms can be further split into two subgroups: the atom that is currently under consideration, termed the bonded-moving atom below; the other atoms in the bonded group are termed bonded-non-moving atoms. A total of three groups of atoms can be classified.

Thus, we can rewrite the covalent-covalent interaction energy as the following four parts.

1) Nonbonded atoms interacting with bonded-non-moving atoms,

$$U = \sum_i^{bonded-non-moving} \sum_j^{nonbonded} (q_i + \vec{\mu}_i \cdot \nabla_i)(q_j + \vec{\mu}_j \cdot \nabla_j) \frac{\text{erf}(\beta_{ij} R_{ij})}{R_{ij}}$$

2) Nonbonded atoms interacting with bonded-moving atom *i*,

$$U = \sum_j^{nonbonded} (q_i + \vec{\mu}_i \cdot \nabla_i)(q_j + \vec{\mu}_j \cdot \nabla_j) \frac{\text{erf}(\beta_{ij} R_{ij})}{R_{ij}}$$

3) Bonded-non-moving atoms interacting with bonded-non-moving atoms,

$$U = \frac{1}{2} \sum_i^{bonded-non-moving} \sum_{j \neq i}^{bonded-non-moving} (q_i + \vec{\mu}_i \cdot \nabla_i)(q_j + \vec{\mu}_j \cdot \nabla_j) \frac{\text{erf}(\beta_{ij} R_{ij})}{R_{ij}}$$

4) Bonded-non-moving atoms interacting with bonded-moving atom *i*,

$$U = \sum_j^{bonded-non-moving} (q_i + \vec{\mu}_i \cdot \nabla_i)(q_j + \vec{\mu}_j \cdot \nabla_j) \frac{\text{erf}(\beta_{ij} R_{ij})}{R_{ij}}$$

Apparently, there should be a fifth part of interaction energy, nonbonded atoms interacting with nonbonded atoms. However, this part of energy does not change in the force calculation, so we omit its expression here. Of course, atom *i*'s self-interaction is also ignored as discussed in the text.

Next force on bonded-moving atom (*i*) can be derived as the negative gradient of the above energy terms. When computing the gradient, we should keep in mind that nothing varies on the nonbonded atoms, only the dipole directions vary on the bonded-non-moving atoms, and both dipole directions and positions of the bonded-moving atoms vary. The above four energy parts thus lead to the following four force components, respectively,

$$(\vec{F}_i)_1 = - \sum_{k}^{bonded-non-moving} \sum_{j}^{nonbonded} \nabla_i(\vec{\mu}_k) \cdot \nabla_k (q_j + \vec{\mu}_j \cdot \nabla_j) \frac{\text{erf}(\beta_{kj} R_{kj})}{R_{kj}}$$

$$(\vec{F}_i)_2 = - \sum_{j}^{nonbonded} \nabla_i(\vec{\mu}_i) \cdot \nabla_i (q_j + \vec{\mu}_j \cdot \nabla_j) \frac{\text{erf}(\beta_{ij} R_{ij})}{R_{ij}}$$

$$- \sum_{j}^{nonbonded} (q_i + \vec{\mu}_i \cdot \nabla_i)(q_j + \vec{\mu}_j \cdot \nabla_j) \nabla_i \frac{\text{erf}(\beta_{ij} R_{ij})}{R_{ij}}$$

$$(\vec{F}_i)_3 = - \sum_{k}^{bonded-non-moving} \sum_{j \neq k}^{bonded-non-moving} \nabla_i(\vec{\mu}_k) \cdot \nabla_k (q_j + \vec{\mu}_j \cdot \nabla_j) \frac{\text{erf}(\beta_{kj} R_{kj})}{R_{kj}}$$

$$(\vec{F}_i)_4 = - \sum_{j}^{bonded-non-moving} \nabla_i(\vec{\mu}_i) \cdot \nabla_i (q_j + \vec{\mu}_j \cdot \nabla_j) \frac{\text{erf}(\beta_{ij} R_{ij})}{R_{ij}}$$

$$- \sum_{j}^{bonded-non-moving} (q_i + \vec{\mu}_i \cdot \nabla_i) \nabla_i(\vec{\mu}_j) \cdot \nabla_j \frac{\text{erf}(\beta_{ij} R_{ij})}{R_{ij}}$$

$$- \sum_{j}^{bonded-non-moving} (q_i + \vec{\mu}_i \cdot \nabla_i)(q_j + \vec{\mu}_j \cdot \nabla_j) \nabla_i \frac{\text{erf}(\beta_{ij} R_{ij})}{R_{ij}}$$

Summing up all four components, the final force expression is,

$$\vec{F}_i = -\sum_{k}^{n} \sum_{j \neq k}^{N} \nabla_i(\vec{\mu}_k) \cdot \nabla_k (q_j + \vec{\mu}_j \cdot \nabla_j) \frac{\text{erf}(\beta_{kj} R_{kj})}{R_{kj}}$$

$$- \sum_{j \neq i}^{N} (q_i + \vec{\mu}_i \cdot \nabla_i)(q_j + \vec{\mu}_j \cdot \nabla_j) \nabla_i \frac{\text{erf}(\beta_{ij} R_{ij})}{R_{ij}}$$

Here, $n$ and $N$ follows the same notation as section 2.3, number of atoms in the bonded group and the system, respectively.

Second, we consider energies caused by induced dipoles. As stated before, the induced energy contains three parts, induced dipoles interacting with covalent multipoles, induced dipoles interacting with induced dipoles, and induced dipole self-energy $\frac{1}{2}\vec{p}\cdot\vec{E}$. From section 2.3, we know that the total induced energy is,

$$\frac{1}{2}\sum_i^N -\vec{p}_i \cdot \vec{E}_i^0$$

where $\vec{E}_i^0$ is the electric field on atom $i$ only by covalent multipoles.

The induced dipoles are determined by the total electric field,

$$\vec{p}_i = \alpha_i \vec{E}_i = \alpha_i (\vec{E}_i^0 - \sum_{j\neq i}^N \vec{\vec{T}}_{ij} \cdot \vec{p}_j)$$

$$\vec{\vec{T}}_{ij} = \nabla_i \nabla_j \frac{\text{erf}(\beta_{ij} R_{ij})}{R_{ij}}$$

Changing the above expressions into the component format, and applying the Einstein's index notation, we obtain

$$p_i^s = \alpha_i(E_i^{0,s} - T_{ij}^{st} p_j^t)$$

Rearrangement leads to

$$\left(\frac{1}{\alpha_j}\delta_{ij}^{st} + T_{ij}^{st}\right) p_j^t = A_{ij}^{st} p_j^t = E_i^{0,s}$$

If we assume $A_{ij}^{st}$ is inversible, and its inverse matrix is $A^{-1}{}_{ij}^{st}$, we have

$$p_i^s = A^{-1}{}_{ij}^{st} E_j^{0,t}$$

$A^{-1}{}_{ij}^{st}$ is a $3N \times 3N$ matrix, symmetrical for both atom index and component index,

$$A^{-1}{}_{ij}^{st} = A^{-1}{}_{ij}^{ts} = A^{-1}{}_{ji}^{st}$$

The gradient of $A^{-1}$ is,

$$\frac{\partial A^{-1}{}_{ij}^{st}}{\partial x_k^w} = -A^{-1}{}_{ii'}^{ss'} \frac{\partial A_{i'j'}^{s't'}}{\partial x_k^w} A^{-1}{}_{i'j}^{s't} = -A^{-1}{}_{ii'}^{ss'} \frac{\partial T_{i'j'}^{s't'}}{\partial x_k^w} A^{-1}{}_{j'j}^{t't}$$

where $w$ refers to coordinate indices $(x^1, x^2, x^3)$. It is obvious that $i'$ or $j'$ has to be equal to $k$ for T to have a nonzero value. We have

$$\frac{\partial T_{i'j'}^{s't'}}{\partial x_k^w} = \frac{\partial T_{kj'}^{s't'}}{\partial x_k^w} + \frac{\partial T_{i'k}^{s't'}}{\partial x_k^w}$$

Based on above relations, the force expressed as the negative gradient of the induced energy is

$$F_i^w = \frac{\partial}{\partial x_i^w}\left(\frac{1}{2}\vec{p}_j \cdot \vec{E}_j^0\right) = \frac{\partial}{\partial x_i^w}\left(\frac{1}{2}A^{-1\,st}_{jk}E_j^{0,s}E_k^{0,t}\right) = \frac{1}{2}\frac{\partial A^{-1\,st}_{jk}}{\partial x_i^w}E_j^{0,s}E_k^{0,t} + A^{-1\,st}_{jk}E_j^{0,s}\frac{\partial E_k^{0,t}}{\partial x_i^w}$$

$$= -\frac{1}{2}A^{-1\,ss'}_{jj'}\frac{\partial T^{s't'}_{j'k'}}{\partial x_i^w}A^{-1\,t't}_{k'k}E_j^{0,s}E_k^{0,t} + p_k^t\frac{\partial E_k^{0,t}}{\partial x_i^w} = -\frac{1}{2}p_{j'}^{s'}\frac{\partial T^{s't'}_{j'k'}}{\partial x_i^w}p_{k'}^{t'} + p_k^t\frac{\partial E_k^{0,t}}{\partial x_i^w}$$

$$= -p_i^{s'}\frac{\partial T^{s't'}_{ik'}}{\partial x_i^w}p_{k'}^{t'} + p_k^t\frac{\partial E_k^{0,t}}{\partial x_i^w}$$

Rewriting the above component format into the vector/tensor format, we have

$$\vec{F}_i = -\sum_{j \neq i}^{N}\vec{p}_i \cdot \nabla_i(\vec{p}_j \cdot \nabla_j)\nabla_i\frac{\text{erf}(\beta_{ij}R_{ij})}{R_{ij}} + \sum_j^N \nabla_i(\vec{E}_j^0) \cdot \vec{p}_j$$

The next step is to evaluate $\nabla_i(\vec{E}_j^0)$. Following the similar strategy used in covalent-covalent interactions, we split the system into two groups: non-moving atoms and moving atom (i.e. atom $i$).

When computing the derivative of field on a non-moving atom $j$, it is worth pointing out that other nonbonded non-moving atoms are not influenced by the virtual displacement of atom $i$, so only bonded non-moving atoms are considered below.

$$\nabla_i(\vec{E}_j^0) = \nabla_i\left(\sum_{k \neq j}^{bonded-non-moving} -\nabla_j(q_k + \vec{\mu}_k \cdot \nabla_k)\frac{\text{erf}(\beta_{kj}R_{kj})}{R_{kj}}\right.$$

$$\left. - \nabla_j(q_i + \vec{\mu}_i \cdot \nabla_i)\frac{\text{erf}(\beta_{ij}R_{ij})}{R_{ij}}\right)$$

$$= \sum_{k \neq j}^{bonded-non-moving} -\nabla_i(\vec{\mu}_k) \cdot \nabla_k\nabla_j\frac{\text{erf}(\beta_{kj}R_{kj})}{R_{kj}} - \nabla_i(\vec{\mu}_i) \cdot \nabla_i\nabla_j\frac{\text{erf}(\beta_{ij}R_{ij})}{R_{ij}}$$

$$- (q_i + \vec{\mu}_i \cdot \nabla_i)\nabla_i\nabla_j\frac{\text{erf}(\beta_{ij}R_{ij})}{R_{ij}}$$

$$= \sum_{k \neq j}^{n} -\nabla_i(\vec{\mu}_k) \cdot \nabla_k\nabla_j\frac{\text{erf}(\beta_{kj}R_{kj})}{R_{kj}} - (q_i + \vec{\mu}_i \cdot \nabla_i)\nabla_i\nabla_j\frac{\text{erf}(\beta_{ij}R_{ij})}{R_{ij}}$$

Here $n$ represents the number of atoms in the bonded group, including atom $i$.

Next we compute the derivative of field on the moving atom, i.e. atom $i$, as follows

$$\nabla_i(\vec{E}_i^0) = \nabla_i \left( \sum_{j \neq i}^{n} -\nabla_i(q_j + \vec{\mu}_j \cdot \nabla_j) \frac{\text{erf}(\beta_{ij}R_{ij})}{R_{ij}} - \sum_{j}^{nonbonded} \nabla_i(q_j + \vec{\mu}_j \cdot \nabla_j) \frac{\text{erf}(\beta_{ij}R_{ij})}{R_{ij}} \right)$$

$$= \sum_{j \neq i}^{n} -\nabla_i(\vec{\mu}_j) \cdot \nabla_j \nabla_i \frac{\text{erf}(\beta_{ij}R_{ij})}{R_{ij}} - \sum_{j \neq i}^{n} (q_j + \vec{\mu}_j \cdot \nabla_j) \nabla_i \nabla_i \frac{\text{erf}(\beta_{ij}R_{ij})}{R_{ij}}$$

$$- \sum_{j \neq i}^{nonbonded} (q_j + \vec{\mu}_j \cdot \nabla_j) \nabla_i \nabla_i \frac{\text{erf}(\beta_{ij}R_{ij})}{R_{ij}}$$

$$= -\sum_{j \neq i}^{n} \nabla_i(\vec{\mu}_j) \cdot \nabla_j \nabla_i \frac{\text{erf}(\beta_{ij}R_{ij})}{R_{ij}} - \sum_{j \neq i}^{N} (q_j + \vec{\mu}_j \cdot \nabla_j) \nabla_i \nabla_i \frac{\text{erf}(\beta_{ij}R_{ij})}{R_{ij}}$$

Here $N$ represents the number of all atoms in the system.